# LeagueBot: A Voice LLM Companion of Cognitive and Emotional Support for Novice Players in Competitive Games

Jungmin Lee

School of Business, Yonsei University, boonddiga00@yonsei.ac.kr

Inhee Cho

Department of Electrical and Electronic Engineering, Yonsei University, choinh@yonsei.ac.kr

Youngjae Yoo[*]

Graduate Program in Cognitive Science, Yonsei University, yooyoungjae@yonsei.ac.kr

Competitive games pose steep learning curves and strong social pressures, often discouraging novice players and limiting sustained engagement. To address these challenges, this study introduces LeagueBot, a large language model-based voice chatbot designed to provide both informational and emotional support during live gameplay in league of legends, one of the most competitive multiplayer online battle arena (MOBA) games. In a within-subjects experiment with 33 novice players, LeagueBot was found to reduce cognitive challenge, performative challenge, and perceived tension. Qualitative analysis further identified three themes: enhanced access to game information, relief from cognitive burden, and practical limitations. Participants noted that LeagueBot offered context-appropriate guidance and emotional support, helping ease the steep learning curve and psychological pressures of competitive gaming. Together, these findings underscore the potential of voice-based LLM companions to assist novice players in competitive environments and highlight their broader applicability for real-time support in other high-pressure contexts.

CCS CONCEPTS • Human-centered computing • Human computer interaction (HCI) • Empirical studies in HCI

**Additional Keywords and Phrases:** Competitive Gaming, Large Language Model, Computational Assistants

## 1 INTRODUCTION

Competitive games are digital games in which multiple players compete under uniform rules and conditions to determine who wins or loses. They represent one of the most commercially successful genres in the gaming market. For example, shooter games generated $7.4 billion in revenue in the PC and console sector in 2024, making them the highest-earning genre [48]. Similarly, the MOBA genre occupies a central position in esports, drawing millions of viewers worldwide through international tournaments and leagues [57]. Esports has developed into a multi-billion-dollar industry, officially included in the Asian Games and even considered for Olympic recognition, an

---

[*] Corresponding Autor

indication of its rapid growth and increasing legitimacy. Other competitive genres, such as real-time strategy (RTS), fighting, and sports games, also rely on players competing for victory [45]. Importantly, it is the tension and sense of accomplishment inherent in competitive play that attract and sustain diverse audiences [28, 58].

Despite their popularity, competitive games are generally difficult to master, and the required skills differ across genres. First-person shooter (FPS) and fighting games demand not only fast reflexes but also instantaneous strategic decision-making [26, 50, 51], requiring sustained practice. Battle royale games pose similar challenges. MOBAs, however, stand out as especially complex. Players must acquire extensive knowledge about champions, items, runes, and skill combinations, in addition to multidimensional strategies such as lane control, objective control, and team coordination. Moreover, because rules and game balance (the "meta") are frequently updated, complexity often increases with time rather than stabilizing. This steep learning curve can overwhelm beginners, leading to frustration, anxiety, and disengagement in high-pressure scenarios [4, 15, 28]. Compounding these challenges, team-based structures mean that individual mistakes directly affect group performance. At the same time, toxic or uncooperative community cultures can intensify social pressure, contributing to player dissatisfaction and ultimately leading to dropout [54].

To address these challenges faced by novice players, the gaming industry has adopted various approaches, including tutorials, practice modes, supportive user interfaces (UIs), and matchmaking systems. For instance, League of Legends (LoL) offers tutorials and practice environments to introduce basic controls and lane concepts. At the same time, Dota 2 offers a guided mode for beginners, and a recommended item build system. Similarly, FPS shooters such as Overwatch feature training grounds for practice controls and hero mechanics, and Heroes of the storm implemented matchmaking policies designed to protect beginners from extreme skill mismatches.

Academic research has likewise identified major challenges in competitive gaming, such as repetitive practice, reliance on prior knowledge, real-time decision-making, and team collaboration, while stressing the importance of providing contextual assistance that does not disrupt gameplay flow. Previous efforts include tutor agents [47], strategy advisory systems utilizing text-to-speech (TTS)/visual cues [13], real-time data visualization [62], and adaptive difficulty management [2, 56]. However, these systems have primarily targeted technical and strategic performance, whereas issues of psychological pressure and emotional support, frequently highlighted in competitive play, have been comparatively neglected.

Meanwhile, large language model (LLM)-based chatbots are increasingly being introduced into games. They have been employed as supportive agents in metaverse environments [24], farming simulations [38], and VR escape rooms [53], where they assist players in understanding tasks, adapting to new contexts, or solving puzzles. Other studies have explored LLMs as creative partners that extend human thinking and reshape teamwork, such as storytelling or board game collaborations [55, 65]. Importantly, LLM-based chatbots have demonstrated the ability to move beyond explaining game rules by interpreting user context and offering emotional support [38]. However, these studies have primarily examined slower-paced environments—narrative-driven RPGs, board games, and metaverse settings—where interactions were mostly text-based and aligned well with the gameplay context.

In contrast, competitive games are fast-paced and require complex controls, making text-based assistance insufficient. Voice interaction, however, has proven effective in similarly demanding contexts such as driving [25]. Recent advances in voice chatbot technology have further enabled natural, conversational interactions, allowing chatbots to act as everyday assistants. For example, studies show that voice agents can support memory through natural conversations [43], enable real-time dialogue augmentation [66], and provide effective feedback in sports training without adding cognitive load [33]. For example, prior studies show that voice agents can improve driving awareness, support memory, and even guide sports training without adding cognitive load, suggesting similar value in competitive gaming. This progress suggests that LLM-based voice agents could also be valuable in competitive gaming, where players face rapid situational changes and heavy information loads.

Nevertheless, research directly examining voice chatbots in competitive game settings remains limited. To address this gap, the present study investigates whether LLM-based voice agents can provide both cognitive and emotional support in competitive gameplay. Specifically, we pose the following research questions:
1) What cognitive and emotional experiences do players encounter during competitive gameplay?
2) What form of support should an LLM-based voice agent provide?
3) What are the potential benefits and limitations of an LLM-based voice agent in competitive gameplay situations?

To explore these research questions, we reviewed prior studies on competitive games. Building on these insights, we developed a prototype LLM-based voice agent and conducted a practical evaluation in a competitive gaming setting.

We analyzed the impact of the LLM-based voice agent on player experience using a within-subjects experimental design and a mixed-methods approach that integrated quantitative experiments with qualitative interviews. This design enabled us to assess not only the effectiveness of the chatbot but also its acceptability and feasibility from multiple perspectives. This study demonstrates the significant value of integrating LLM-based voice agents into competitive games to enhance player experience. It also identifies key design considerations for applying such agents in fast-paced, high-pressure environments, offering implications for both research and industry.

## 2 RELATED WORKS

### 2.1 Player experience in competitive games

Competitive games are structured around multiple players competing against one another, often in team-based formats [18]. Players strive for victory by collaborating with teammates while competing against opponents [17, 45]. By design, such games present steep learning curves [28, 50, 51]. Yet, overcoming these challenges and developing skills can yield strong feelings of achievement and satisfaction [19, 28].

However, the competitive intensity and complex mechanics of these games can create unfavorable experiences for novices. Gee (2003) [19] described successful games as "pleasantly frustrating," where challenges remain demanding but achievable, thereby fostering learning and immersion. Similarly, Csikszentmihalyi et al.'s (2014) [11] Flow Theory posits that immersion arises when task difficulty is balanced with a player's ability. While competition enhances enjoyment, for beginners it often produces skill gaps and excessive difficulty, leading to negative experiences [40, 41]. To address this, competitive games have long sought ways to tailor experiences to players' skill levels [2, 26, 62]. These issues are particularly acute in MOBA games, which generate higher frustration and challenge than many other genres [28].

Team-based play further compounds the challenges faced by novices. Adinolf and Turkay (2018) [1] examined toxic behavior in competitive games, noting that competition often fuels hostility that undermines collaboration. Similarly, Grandprey-Shores et al. [22] found that toxic behaviors such as blaming teammates intensify as competitiveness increases. Toxic behavior is especially prevalent in MOBAs, where teamwork and social interaction coexist with extreme competitiveness. Tyack et al. (2016) [58] observed that in MOBAs, players frequently perceive teammates as task-oriented instruments rather than social partners, particularly when playing with strangers. Kou (2020) [34] showed in LoL that competitiveness, intra-team conflict, and perceptions of defeat are primary triggers of toxic behaviors. Novice players are often perceived as underperformers, which can exacerbate frustration, conflict, and hostility, ultimately creating an unwelcoming environment for newcomers. [1, 34, 36].

Thus, the steep learning curve and competitive dynamics of these games not only hinder individual progress but also strain group interactions. Paradoxically, these same challenges enable players to achieve mastery and experience collective accomplishment, offering deep satisfaction and motivation. In this sense, competitive gaming embodies a duality of frustration and fulfillment, as well as cooperation and conflict.

### 2.2 Psychological burden of competitive games

Competitive games are structured around competitive systems [17], which exert continuous psychological pressure on players. Because outcomes are clearly defined as victory or defeat, and because individual errors in team-based settings can directly cause collective losses, players often experience persistent tension during matches. Repeated defeats and intra-team conflicts amplify frustration, which may further impair performance and trigger a destructive cycle commonly referred to as "tilt." These pressures are intensified by factors such as skill level disparities and the tendency to blame teammates for mistakes [34, 63].

To cope with these negative experiences, players adopt diverse emotional regulation strategies. Kou and Gui (2020) [35] observed that LoL players under stress often set smaller goals, mute disruptive teammates, listen to music, or reframe the experience by reminding themselves that "it's just a game." Beres et al. (2023) [5] categorized such strategies into five types: situation selection (e.g., taking breaks), situation modification (e.g., blocking chat), attentional deployment, cognitive reappraisal, and response suppression. These findings suggest that players actively manage, rather than merely endure, psychological pressure. Nonetheless, current game systems rarely facilitate these coping strategies, leaving many players to disengage temporarily as a means of relief [35].

Importantly, emotional regulation is not confined to the individual but is also shaped by social dynamics. Kou et al. (2020) [35] and Seim et al. (2025) [54] demonstrated that experiences of winning or losing streaks are influenced by social factors such as camaraderie, guilt, and peer expectations, all of which affect pressure levels and disengagement tendencies. Wu et al. (2021) [63] further highlighted, in a study of aspiring professional players, that teammate conflict is the most significant stressor, requiring both personal regulation and team-level interventions, such as improved communication and training in growth-oriented mindsets.

Playing with friends has been shown to mitigate competitive stress. Tyack et al. (2016) [58] found that friend-based play fosters camaraderie and emotional support, reducing stress and enhancing vitality, whereas playing with strangers intensifies outcome-oriented attitudes and negative experiences. In summary, competitive players employ both individual and social strategies to manage stress, with cooperative play serving as a central protective factor that alleviates psychological burdens and promotes more positive engagement.

### 2.3 Computational tools for supporting competitive games

Scholars have explored various computational tools to support players in MOBA games and related genres. Kleinman et al. (2022) [32] identified four major challenges: team collaboration, knowing what to do next, monitoring the game state, and keeping track of skill mastery. These challenges arise not only from insufficient technical skills but from the fast-paced and complex decision-making environments of competitive games. Kleinman et al. (2022) [31], emphasized the importance of a context-specific and non-intrusive support system that does not disrupt gameplay flow.

Several prototypes illustrate this approach. do Nascimento Silva and Chaimowicz (2015) [47] created a tutoring agent using the LoL champion Soraka, which provided novices with warnings and advice via smart pings when their health was low or when they faced imminent danger. Similarly, Cunha et al. (2015) [13] developed RTS games, which used visual cues and TTS audio to assist resource management and unit production, thereby improving performance. In the context of fighting games, Ibrahim et al. (2023) [26] implemented visual aids and context-sensitive tutorials to guide novice players, further demonstrating the potential of computational support tools in facilitating learning and reducing entry barriers.

Recent studies have explored the use of real-time data visualization and external devices to assist players. Rijnders et al. (2022) [52], in a study of LoL players, confirmed the effectiveness of such systems and identified six design principles for effective feedback: information density, variation in information value, visual summarization, situational awareness, explanation of reasoning, and adaptation to skill level. Similarly, Wang et al. (2024) [62] analyzed 15 commercial companion tools for LoL and Valorant, showing that these systems provide not only pre- and post-match assistance but also in-game features such as real-time statistics, event-driven tips, champion selection support, and monster respawn timers. Building on this, Wang et al. (2025) [61] emphasized the importance of personalization and transparency, noting that onboarding-oriented, simplified support benefits novices, while advanced analytics are more suitable for experienced players.

In summary, support tools for competitive games have evolved to deliver immediate, contextual, and skill-sensitive feedback, thereby reducing player challenges. Yet, despite evidence that competitive play imposes intense psychological pressure [4, 37] and considerably shapes the gaming experience [4], most computational tools remain narrowly focused on technical and strategic functions. Tools that integrate emotional support remain scarce, leaving a critical research gap in supporting players of MOBA games.

**2.4 Large language models as game companions**

Recent research has begun integrating large language models (LLMs) into gaming, expanding their role from simple information delivery to enhancing player experiences. For example, Zhu et al. (2023) [65] utilized an LLM as a dungeon master assistant in Dungeons & Dragons, aiding in the interpretation of rules and the generation of narratives. Rather than replacing storytelling, the AI served as a creative partner. Likewise, Sidji et al. (2024) [55] utilized an LLM agent in Codenames, demonstrating that it altered team dynamics and served as an equalizer for less experienced players. Hong (2024) [24] found that in metaverse onboarding, LLM-powered non-player characters (NPCs) facilitated learning through context-aware support and increased immersion. Similarly, Lee et al. (2025) [38] demonstrated in Stardew Valley that chatbots could deliver both beginner-friendly guidance and emotional engagement, reinforcing the potential of LLM companions as relational game partners.

Most prior studies, however, focused on slower-paced environments such as narrative-driven RPGs, board games, and metaverse platforms, where text-based interaction is sufficient. In contrast, fast-paced competitive games demand immediate responses, making text alone inadequate. Advances in voice interfaces now enable real-time support without requiring players to shift screens. For example, Huang et al. (2024) [25] argued that dynamic LLM-based voice assistants improve awareness and driving performance. Maniar et al. (2025) [43] developed a voice-based support service for older adults with memory limitations, enabling natural and scalable conversations. Zulfikar et al. (2024) [66] combined LLMs with text-to-speech technology for real-time memory augmentation in complex dialogues, while Ko et al. (2025) [33] demonstrated that LLM-based verbal feedback in augmented-reality (AR) sports training, such as golf swing practice, was effective without increasing cognitive load.

These findings suggest that LLM companions can be applied to fast-paced genres. Nevertheless, little research has examined the effectiveness, acceptance, and feasibility of voice-based LLM interactions in dynamic competitive games. To address this gap, the present study investigates the cognitive and emotional support offered by a voice-based LLM companion, LeagueBot, for novice LoL players and explores its design implications.

## 3 DEVELOPMENT OF LEAGUEBOT

### 3.1 Selection of a competitive game

This study focuses on MOBA games. Compared to other genres, MOBAs are particularly prone to inducing frustration and stress due to their highly competitive structure [28]. LoL, developed by Riot Games, is one of the most successful MOBAs worldwide. Its most popular mode is a 5v5 battle on Summoner's Rift, where teams aim to destroy the opposing Nexus. Players progress through phases of laning in the top, jungle, mid, and bottom lanes, competing for objectives, and eventually engaging in large-scale team fights. Success requires consideration of numerous strategic elements, including lane control, objective timing, and positional coordination.

LoL has an especially steep learning curve. At launch, the game offered only 17 champions, but as of September 2025, it includes 171, each with distinct abilities and unique gameplay. Players must understand not only their own champion but also the skills of their teammates and opponents. Each champion requires a unique item build, and players must master item interactions to optimize their performance. Beyond this knowledge, advanced strategies, such as lane dynamics, coordinated team fighting, and objective control, require additional mastery. Frequent updates and seasonal overhauls further complicate gameplay, as rebalancing forces players to continually adapt.

To ease the onboarding process, LoL provides several practice modes. Players are restricted from ranked matches, which affects their competitive standings, until they reach level 30. Before that, they can gain experience through bot matches (Introductory, beginner, or intermediate) or through unranked matches with other players. Ranked games, unlocked after level 30, feature greater competitive intensity despite matchmaking systems designed to balance skill levels. Newcomers often face steep barriers to entry and considerable psychological pressure in these environments. Conflicts among teammates are common and can make adaptation difficult, sometimes discouraging continued play.

Given these characteristics, LoL exemplifies a steep learning curve, heightened competitiveness, and strong social pressures associated with team-based play. It thus provides an ideal environment to evaluate systems intended to reduce or mitigate such burdens.

### 3.2 Initial design

*3.2.1 Design goals*

The primary goal of the LeagueBot is to improve the experience of novice LoL players. The system was designed with three objectives. First, it supports adaptation to the fast-paced game environment through real-time interaction. To achieve this, the LeagueBot was implemented as a voice-based chatbot capable of interpreting and responding to in-game situations. Second, it facilitates access to complex knowledge by efficiently communicating information on champion abilities, item builds, and strategic decisions. Ultimately, it provides emotional support and companionship, offering comfort and empathy similar to interacting with a friend. By fostering relational engagement, the LeagueBot helps alleviate the psychological pressure often experienced by new players.

*3.2.2 Implementation of LLM-based voice chatbot*

Voice-based interaction was implemented using ElevenLabs' conversational AI, which offers a range of synthetic voices and tools for building LLM-powered assistants. Because accessing in-game data from LoL required its API, the system was developed as a desktop application with electron. The chatbot was integrated through ElevenLabs' JavaScript SDK, while GPT-4.1, the most recent model available at the time, served as the underlying LLM to ensure reliable, up-to-date responses. To accommodate the primary user base, a male, Korean-supported voice was selected. This design choice enhanced familiarity and immersion during gameplay.

*3.2.3 Chatbot persona*

The chatbot persona was created using prompt engineering based on ElevenLabs' design framework, which recommends six key elements: personality, environment, tone, goal, guardrails, and tools. These elements ensured consistent interaction patterns and natural voice delivery. For information delivery, such as strategic suggestions, explanations of game mechanics, and situational advice, the design drew on esports coaching methodologies identified by Lee, H. et al. (2025) [37]. Personality and tone were shaped by prior studies that showed LoL players frequently engage with friends and that social contexts enhance enjoyment [58]. Accordingly, the chatbot was designed to act like a supportive teammate, offering friendly, cooperative interactions. The initial prototype was limited to in-game use only, reflecting the real-time, context-sensitive nature of MOBAs, where immediate guidance both improves effectiveness and helps alleviate in-match psychological stress.

*3.2.4 Game information and context awareness*

Because the LLM lacked up-to-date knowledge of LoL, supplementary documents were provided, including details about new champions (e.g., Yunara in Korean) and items such as Atakan. To enable context-aware responses, LeagueBot used ElevenLabs' contextual update and tools functions. Contextual updates provided one-time information during transitions, such as from champion selection to match start, including details of both teams' champion choices. The tools function, leveraging the LLM's function-calling capability, was configured to fetch additional data on demand.

Once gameplay began, the system continuously retrieved real-time data through the LoL API, which the chatbot applied in two ways:
- Game data: scoreboard, match progress, and primary objective.
- User data: runes, skills, statistics, item builds, and current gold.

Game data allowed the chatbot to evaluate match flow and relative team strength, supporting advice on contesting objectives or initiating fights. User data provided personalized guidance, such as optimal skill usage, item builds, or recommendations on whether to farm or engage in combat. By integrating global and individual perspectives, the chatbot offered context-sensitive, adaptive support throughout gameplay.

### 3.3 Pilot Test and Final Design

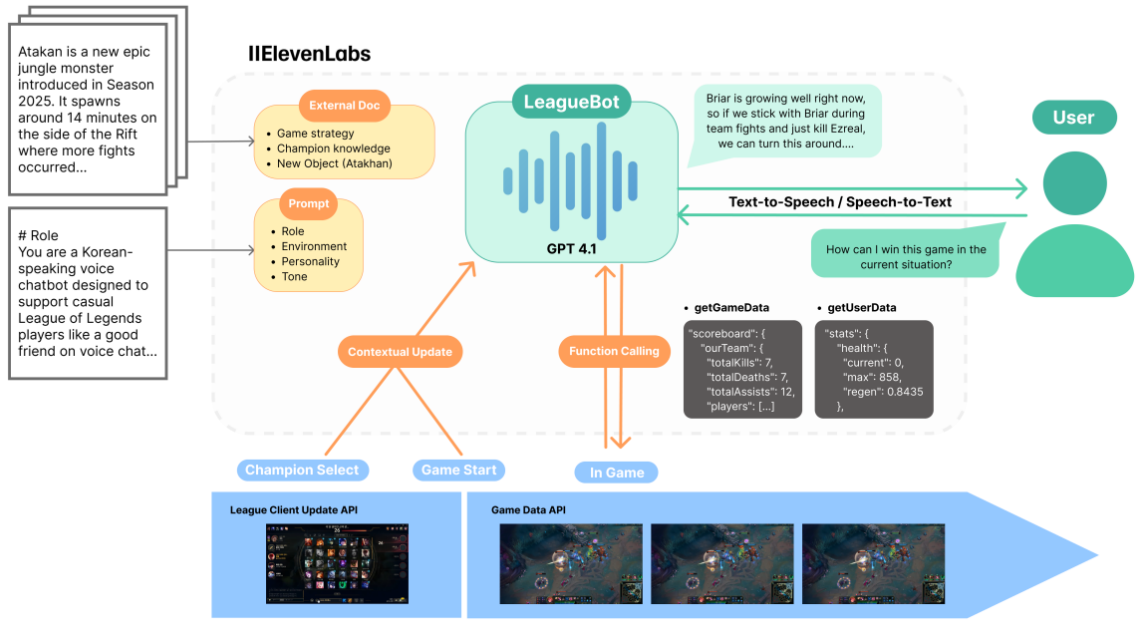

Figure 1. The overall framework of LeagueBot

*3.3.1 Participants and procedure*

This pilot study aimed to examine the detailed design features of the chatbot and identify potential issues and areas for improvement during live gameplay. Although the primary focus was on novice players, participants with different levels of expertise were included to capture a broad range of feedback on chatbot interactions.

Six participants took part in the study: four women and two men, aged 23 to 29. Their experience with LoL varied considerably, ranging from long-term players with more than six years of consistent engagement to beginners with less than a year of experience, and even to occasional users who had played only a few matches. Player levels ranged from 1 to 262. At the same time, novice users were included in the pilot phase because they are typically more focused on immersion and may find it challenging to provide detailed feedback. In contrast, experienced players are more likely to identify specific functional limitations of the chatbot.

The test was carried out on laboratory computers. Participants first received a short introduction to the chatbot's functions and operation. They were then given approximately five minutes in practice mode to interact with the chatbot and become familiar with the interface. The main session consisted of one full game played with the chatbot, followed immediately by a semi-structured interview of captured perceptions of gameplay and chatbot use. Interviews were guided by questions designed to evaluate the chatbot's performance from multiple perspectives:
Functional suitability:
- "Compared with your previous gameplay experience, how was playing with the chatbot?"
- "Were there any cases where the chatbot failed to answer your questions?"

Usability:

- "Did you experience any discomfort while conversing with the chatbot?"
- "Did the chatbot interfere with your gameplay?"

Timing and contextual appropriateness:
- "How was the timing of the chatbot's advice?"
- "Did you feel the chatbot accurately understood the game situation and provided appropriate advice?"
- "Were there moments when the chatbot distracted you from focusing on the game?"

Interaction quality:
- "How did you perceive the chatbot's personality?"
- "What impression did you have of the chatbot's speaking style and voice?"

*3.3.2 Feedback and implementation*

Participants generally provided positive feedback on the chatbot's ability to deliver information and convey emotional responses. However, several areas for improvement were identified during the pilot test (Table 1).

First, unsolicited interactions were highlighted as disruptive. The feature that allowed the chatbot to initiate conversation after a period of silence often interrupted moments requiring concentration, such as battles. For example, one participant noted, "When I was focused on fighting, the chatbot suddenly spoke and startled me," while another remarked, "I wish it wouldn't talk when I'm concentrating." To resolve this issue, we turned off the proactive speech feature in ElevenLabs, ensuring that interactions occurred only in response to user input.

Second, participants reported a lack of clarity in information delivery. The coaching-inspired conversational style, which encouraged players to make independent decisions, was confusing. Instead, participants preferred concise and direct guidance. For instance, when a player asked which item to purchase, the chatbot replied, "Both are fine," which hindered quick decision-making. Moreover, explanations that listed multiple items sequentially forced players to wait until the end, making them impractical for fast-paced contexts. To address this, we modified prompts to provide information in a front-loaded, concise manner, presenting the primary recommendation first, followed by optional elaboration if requested.

Third, repetition and unnaturalness in emotional reassurance were observed. The frequent use of identical or similar comfort phrases reduced the naturalness of interactions. To enhance variability, we increased the LLM's temperature setting from 0 to 0.5, allowing for more diverse and natural responses.

Fourth, expressed different preferences for voice options. Some female players suggested that a female voice would feel more relatable and engaging. In response, we added female voice options, enabling users to customize their preferred voice for a more personalized experience.

Finally, insufficient support was noted during the selection phase of the champion. While some participants desired assistance at this stage, the prototype limited chatbot interactions to in-game contexts. To address this limitation, we extended functionality to include champion selection. The system was revised to allow conversations to begin during champion select, with contextual updates refreshing team and opponent information each time selections were finalized.

By incorporating these improvements, LeagueBot became more adaptable, improved clarity in information delivery, offered greater emotional variability, and supported personalized interaction experiences. The finalized design of LeagueBot is shown in Figure 1, and the prompt used in the study is listed in Appendix 1.

Table 1 Summary of interview findings and corresponding design insights

| Dimension | Participant Responses (Examples) | Design Insight |
| --- | --- | --- |
| Functional suitability | Guidance was vague | Provide clear, front-loaded recommendations; avoid ambiguous answers. |
| Usability | Unsolicited speech during fights | Disable proactive speech; ensure responses are user-initiated. |
| Timing & context | Explanations too long/sequential | Adopt concise, context-sensitive messaging; primary suggestion first, optional detail later. |
| Interaction quality | Reassurance repetitive; limited voice options | Increase response variability; Offer customizable voice options. |
| Additional feedback | No support in champion selection | Extend functionality to champion select phase with contextual updates. |

## 4 METHOD

### 4.1 Study design

This study adopted a mixed-methods design that integrated quantitative experiments with qualitative interviews. The experimental phase used a within-subjects design, allowing each participant to experience both conditions: (1) Playing the game with LeagueBot, and (2) Playing without the chatbot. The conditions were randomized to reduce potential carryover effects, such as learning or fatigue. After completing the experiment, participants were interviewed to explore their experiences in greater depth, thereby complementing the quantitative findings and providing a deeper understanding of the context. The experimental setting was designed to mirror participants' typical LoL play environments, ensuring ecological validity. Participants were allowed to use their usual in-game configurations to maintain a natural playing style. Supplementary activities, such as consulting champion guides or item builds, were also allowed. Because LoL's in-game text chat often contains toxic language and profanity, chat functionality was turned off during gameplay. This measure was implemented to prevent adverse psychological effects and to isolate the chatbot's influence on users.

### 4.2 Participants

Participants were recruited among novice LoL players through both online (community boards and social media) and offline (posters on university boards) channels in Seoul. A pre-survey screening process followed recruitment. A novice was defined as a player who had not yet reached level 30 in LoL (and was therefore ineligible for ranked games) or who, despite reaching level 30, had played no more than 20 ranked matches. This classification was adopted because ranked play requires higher competitiveness and greater complexity, making limited ranked experience a reasonable indicator of novice status.

The pre-survey also gathered background information, including the frequency of play, game level, year participants began playing, orientation towards competitive play (on a 7-point scale), preferred game modes (e.g., bot match, normal game, or all random all mid (ARAM), and frequency of LLM use.

A total of 34 participants met the criteria and were included in the study. Their average age was 23.4 years (SD = 2.5; range, 20–29), comprising 21 men and 13 women. Participants' current game levels ranged from 1 to 189, with a mean of 51.7 (SD = 45.3). Start years ranged from 2013 to 2025, with 14 participants (41.1%) having begun playing since 2023 or later, thus classified as recent novices. Regarding gameplay frequency, most participants (47.1%) played infrequently (once a month or less), while 23.5% played 1-3 times per week. In terms of preferred modes, 44.1% mainly played ARAM, while 26.4% preferred normal (blind or draft) games. Participants showed a low to moderate competitive orientation, with a mean score of 2.4 (SD = 1.5) on a 7-point scale. Regarding the frequency of generative AI usage, 41.2% of the respondents used AI tools daily, 26.5% used them 1-3 times per week, and 20.6% used them 4-6 times per week. One participant was excluded from the final analysis after being identified as a Bronze-tier player in the current season, indicating sufficient ranked experience. All participants received an incentive of 30,000 KRW (approximately $21.50). Detailed participant information is provided in Appendix 2.

### 4.3 Procedure

Eligible participants were individually contacted, informed about the study's purpose, duration, and data privacy measures, and then scheduled for experiment sessions. Upon arrival at the laboratory, participants were briefed on the procedures and provided written informed consent. Each participant completed two game sessions, each consisting of one normal LoL match lasting approximately 20–30 minutes. Including setup and pre- and post-game surveys, each session lasted approximately 60–70 minutes.

Control Condition: (Without Chatbot). Participants played a standard LoL match under typical conditions. Before the main session, they received approximately five minutes of practice time in warm-up mode. During the game, participants were free to select their champion, rune configuration, and summoner spells. To replicate a natural gaming environment, they were also permitted to consult online resources, such as champion guides, item builds, or strategic advice, as needed.

Experimental condition (With chatbot). In this condition, participants played with the LeagueBot activated. The chatbot was accessible from the champion selection phase and remained available throughout the match. As in the control condition, participants completed a five-minute warm-up in practice mode, during which they could already interact with the chatbot. They were free to request advice regarding champion abilities, itemization, or strategic decisions. The chatbot provided context-sensitive guidance using real-time game data (e.g., kill score, champion attributes, and item builds) retrieved through the game's API. All interactions were automatically logged, with participants informed of the recording process and their consent obtained before participation. The dialogue data were later analyzed to examine both the nature of participants' queries and the contextual factors surrounding their interactions with LeagueBot.

### 4.4 Data collection

*4.4.1 Measures*

To examine how a voice-based LLM chatbot affects the cognitive and psychological burdens of novice MOBA players, participants' perceived challenge was assessed.

Perceived challenge (30 items): The challenge originating from recent gameplay interaction scale (CORGIS) [15] was employed to evaluate the level of challenge experienced during gameplay. Although CORGIS comprises four dimensions, cognitive, emotional, performative, and decision-making, this study used only three, excluding the emotional dimension. Items related to emotional challenge were removed because they primarily address narrative immersion and moral dilemmas, which are less relevant to LoL, a competition-oriented multiplayer game.

To investigate how the chatbot influences the overall game experience of novice MOBA players, player experience was measured using the following indicators:

Intrinsic motivation inventory (IMI; 12 items): A shortened version of the interest/enjoyment and pressure/tension subscales [44] was administered to assess emotional experiences.

1) Interest/Enjoyment measured intrinsic motivation and overall enjoyment during gameplay.

2) Pressure/Tension captured the psychological pressure participants experienced.

These metrics enabled analysis of how the chatbot influenced both enjoyment and tension level.

Chatbot usability (10 items): The system usability scale (SUS) [8] was used to assess the chatbot's usability. This concise and robust tool comprises 10 items, each rated on a 5-point Likert scale, yielding a total score ranging from 0 to 100. Scores of 68 represent the average benchmark, 70 or higher indicates above-average usability, and 80 or higher reflects excellent usability. Given its reliability and simplicity, SUS was considered well-suited for evaluating usability in the real-time, interactive setting of LeagueBot.

*4.4.2 Interviews*

Upon completing all gameplay sessions and surveys, participants engaged in semi-structured interviews lasting approximately 20–30 minutes. With informed consent, all sessions were audio-recorded and subsequently transcribed for qualitative analysis. The interviews addressed five themes.

1. Participants' experiences and impressions of gameplay with and without LeagueBot.
2. Their perceptions of LeagueBot as an entity.
3. Comparisons between playing with LeaguBot and playing with a human teammate, with attention to how LeaguBot influenced the gaming experience.
4. Reflections on the advantages and disadvantages of using the chatbot relative to traditional information sources, as well as the features that distinguish LeagueBot from other AI services.
5. Overall impressions and suggestions for improving LeagueBot.

*4.4.3 Game data*

To complement the qualitative findings from the interviews and to better understand the overall usage context, additional game data were collected. In the chatbot condition, all conversation logs between participants and LeagueBot were recorded with timestamps. Additionally, all gameplay sessions were screen-recorded to capture the contextual details of how interactions with LeagueBot unfolded during actual matches. These data, combined with the interview results, were analyzed to provide a more comprehensive understanding of participants' experiences.

**4.5 Data analysis**

For the quantitative data, because the same participants experienced both the chatbot and control conditions, paired t-tests were conducted to examine mean differences between the two conditions across the subscales of the measurement instruments (CORGIS, IMI). All analyses were performed using IBM SPSS (version 31).

For the qualitative data, a thematic analysis was conducted on the interview materials [7]. First, the interview recordings were transcribed using Clova Note, after which the research team manually corrected transcription errors. All authors then repeatedly reviewed the transcripts to ensure thorough familiarity with the content. The first author conducted initial coding, followed by iterative workshops with all authors to reach a consensus on the final themes.

## 5 Results

### 5.1 Measurement validation

Confirmatory factor analysis (CFA) and reliability testing were conducted to assess the validity and reliability of the measurement instruments. The instruments comprised five constructs: cognitive challenge (COGCHAL), performative challenge (PERCHAL), decision-making challenge (DECCHAL), enjoyment (ENJOY), and tension (TENSION).

To evaluate convergent validity, a CFA was performed using principal component analysis with varimax rotation and Kaiser normalization. The study resulted in the removal of four COGCHAL items, five DECCHAL items, and one ENJOY item due to low factor loadings. The retrained items all exhibited loadings greater than .6, which is considered acceptable [23]. Furthermore, the average variance extracted (AVE) values exceeded .5, supporting convergent validity.

Reliability was assessed using composite reliability (CR), and Cronbach's α was calculated. All constructs demonstrated CR values above .6 and Cronbach's α coefficients exceeding .7, confirming satisfactory internal consistency [9, 16].

Overall, the five constructs exhibited adequate convergent validity and reliability. The detailed results are presented in Table 2.

Table 2. Results of Measurement Validation

| Construct | Item | Factor loading | AVE | Composite reliability | Cronbach's alpha |
|---|---|---|---|---|---|
| COGCHAL | COGCHAL1 | 0.816 | 0.604 | 0.914 | 0.922 |
|  | COGCHAL2 | 0.852 |  |  |  |
|  | COGCHAL3 | 0.809 |  |  |  |
|  | COGCHAL4 | 0.757 |  |  |  |
|  | COGCHAL7 | 0.706 |  |  |  |
|  | COGCHAL8 | 0.793 |  |  |  |
|  | COGCHAL9 | 0.695 |  |  |  |
| PERCHAL | PERCHAL1 | 0.889 | 0.654 | 0.902 | 0.891 |
|  | PERCHAL2 | 0.910 |  |  |  |
|  | PERCHAL3 | 0.603 |  |  |  |
|  | PERCHAL4 | 0.764 |  |  |  |
|  | PERCHAL5 | 0.838 |  |  |  |
| DECCHAL | DECCHAL1 | 0.723 | 0.509 | 0.675 | 0.878 |
|  | DECCHAL2 | 0.704 |  |  |  |
| ENJOY | ENJOY1 | 0.748 | 0.629 | 0.909 | 0.909 |
|  | ENJOY2 | 0.825 |  |  |  |
|  | ENJOY3 | 0.719 |  |  |  |
|  | ENJOY4 | 0.814 |  |  |  |
|  | ENJOY5 | 0.838 |  |  |  |
|  | ENJOY6 | 0.874 |  |  |  |
| TENSION | TENSION1 | 0.718 | 0.512 | 0.842 | 0.832 |
|  | TENSION2 | 0.723 |  |  |  |
|  | TENSION3 | 0.715 |  |  |  |
|  | TENSION4 | 0.712 |  |  |  |
|  | TENSION5 | 0.761 |  |  |  |

### 5.2 Descriptive statistics

A total of 33 participants played games under both the chatbot and non-chatbot conditions. We first report descriptive statistics for the dependent variables, game outcome data, and chatbot usage data. Game outcome data include win rate, game duration, and the KDA ratio (Kills + Assists to Deaths). KDA is a commonly used metric for evaluating performance in LoL [46, 49], with higher values indicating better performance. When the number of deaths was zero, a value of one was substituted, consistent with common practice in LoL research [46]. Chatbot usage data includes the total number of conversations with the chatbot and the SUS score. Detailed results are summarized in Table 1.

For the dependent variables, in the chatbot condition, cognitive challenge was M = 4.58 (SD = 1.32), whereas in the non-chatbot condition, it was M = 5.26 (SD = 1.22). Performative challenge was M = 5.50 (SD = 1.16) in the chatbot condition and M = 6.08 (SD = 0.69) in the non-chatbot condition. Enjoyment was M = 5.21 (SD = 1.11) in the chatbot condition and M = 5.13 (SD = 1.01) in the non-chatbot condition. Tension was M = 3.61 (SD = 1.26) with the chatbot and M = 4.08 (SD = 1.25) without it. Finally, the decision challenge was M = 4.44 (SD = 2.02) in the chatbot condition and M = 4.50 (SD = 1.82) in the non-chatbot condition.

Regarding game outcome data, the chatbot condition showed a mean win rate of M = 0.47 (SD = 0.51), an average game duration of M = 25.88 minutes (SD = 7.90), and a mean KDA of M = 3.47 (SD = 3.80). In the non-chatbot condition, the mean win rate was M = 0.56 (SD = 0.50), the average game duration was M = 26.81 minutes (SD = 8.76), and the mean KDA was M = 2.54 (SD = 2.72).

Additionally, the average length of chatbot conversation was M = 66.03 (SD = 31.65), and the SUS score for LeagueBot was 76.7, indicating above-average usability [3].

Table 3. Descriptive Statistics

|  | Chatbot Mean (SD) | No Chat Mean (SD) | p-value |
|---|---|---|---|
| **Dependent Variable** | | | |
| Cognitive challenge | 4.58 (1.32) | 5.26 (1.22) | 0.002 |
| Performative challenge | 5.50 (1.16) | 6.08 (0.69) | 0.003 |
| Decision-making challenge | 4.44 (2.02) | 4.50 (1.82) | 0.862 |
| Enjoyment | 5.21 (1.11) | 5.13 (1.01) | 0.724 |
| Tension | 3.61 (1.26) | 4.08 (1.25) | 0.043 |
| **Game results** | | | |
| Win rate | 0.47 | 0.56 | 0.620 |
| Game time (min) | 25.88 (7.90) | 26.81 (8.76) | 0.517 |
| KDA ratio | 3.47 (3.80) | 2.54 (2.72) | 0.214 |
| **Chatbot usage** | | | |
| Conversation turns | 66.03 (31.65) | — | — |
| System usability score | 76.7 | — | — |

## 5.3 Results of gameplay

A comparison of win rates between the chatbot condition (45%) and the non-chatbot condition (55%) revealed no significant difference, $\chi^2(1) = 0.24$, p = .62. Thus, the presence of the chatbot did not influence winning outcomes. Regarding game duration, the chatbot condition (M = 1528.82 s, SD = 431.07) averaged 73.67 seconds shorter than the non-chatbot condition (M = 1602.48 s, SD = 527.55). However, this difference was not significant, t(32) = -0.66, p = .517, Cohen's d = -0.11, (small effect) [10], indicating a negligible effect. Similarly, participants in the chatbot condition (M = 3.47, SD = 3.80) achieved a kill/death assist (KDA) ratio 0.93 points higher than those in the non-chatbot condition (M = 2.54, SD = 2.72). Still, the difference was not statistically significant, t(32) = 1.27, p = .214. Overall, the chatbot did not substantially influence win rates, game duration, or combat performance. Overall, the chatbot did not substantially influence win rates, game duration, or combat performance. Importantly, the absence of significant gameplay differences between conditions ensures that any observed effects on the dependent variables can be attributed to the chatbot itself rather than disparities in game outcomes.

## 5.4 Effectiveness of LeagueBot

This experiment examined the effects of chat condition (chat vs. no chat) on cognitive, performative, and decision-making challenges (CORGIS), as well as enjoyment and tension (IMI). Paired-sample t-tests with two-tailed hypotheses were conducted, and results are reported with descriptive statistics, mean difference, 95% confidence interval, t-value, p-value, and effect size (Cohen's d). Following Cohen's (1988) [10] guidelines, with d = 0.2 considered small, d = 0.5 medium, and d = 0.8 large.

*Cognitive challenge.* Participants reported considerably lower scores in the chatbot condition (M = 4.58, SD = 1.32) than in the non-chatbot condition (M = 5.26, SD = 1.22), t(32) = 3.35, p = .002. The mean difference was 0.68, corresponding to a minimum effect size (d = 0.58), suggesting that the chatbot reduced cognitive load.

*Performative challenge.* scores were also considerably lower in the chatbot condition (M = 5.50, SD = 1.16) compared with non-chatbot condition (M = 6.08, SD = 0.69), t(32) = 3.19, p = .003. The mean difference of 0.59 reflected a medium effect size (d = 0.56), indicating that the chatbot eased performance-related demands.

*Decision-making challenge.* No significant difference was found between the chatbot (M = 4.44, SD = 2.02) and non-chatbot condition (M = 4.50, SD = 1.82), t(32) = 0.18, p = .862, d = 0,03. This result suggests that the chatbot did not influence the decision-making burden.

*Enjoyment.* The chatbot condition (M = 5.21, SD = 1.11) and the non-chatbot condition (M = 5.13, SD = 1.01) did not differ substantially, t(32) = -0.36, p = .724, d = -0.06, indicating negligible effects on enjoyment.

*Tension.* Participants reported considerably lower tension in the chatbot condition (M = 3.61, SD = 1.26) compared with the non-chatbot condition (M = 4.08, SD = 1.25), t(32) = 2.11, p = .043. The mean difference of 0.47 reflected a small to medium effect (d = 0.37), indicating that the chatbot had a moderate effect in reducing tension.

*Summary.* The chatbot substantially reduced cognitive and performative challenges and lowered tension, but it did not affect decision-making challenges or enjoyment (Figure 2).

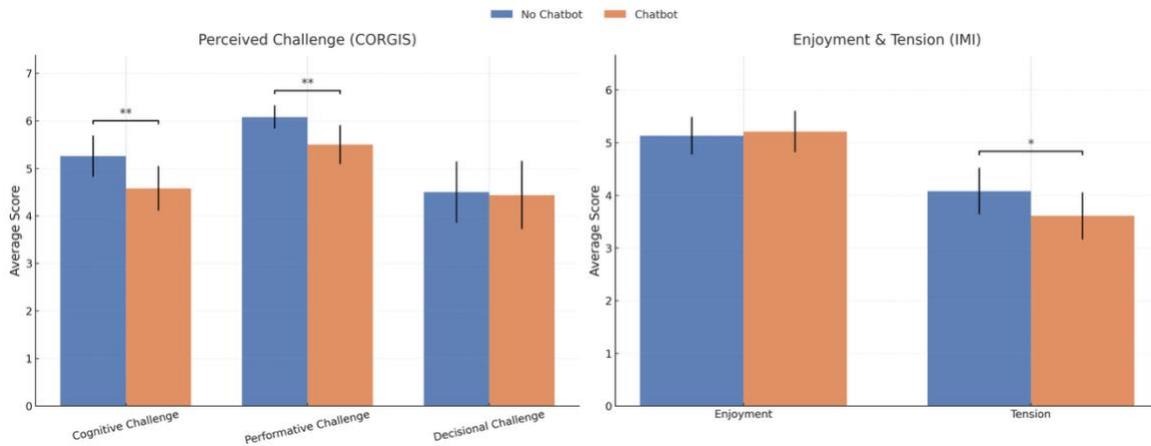

Figure 2. Mean scores of perceived challenge (CORGIS), enjoyment, and tension (IMI) across conditions. Error bars indicate 95% confidence intervals. Significance levels are denoted as follows: p < .05 (*), p < .01 (**).

### 5.5 Results of thematic analysis

The thematic analysis revealed three overarching themes: (1) advantages of acquiring game-related information, (2) burden alleviation through the chatbot, and (3) pros and cons of real-world implementations. As shown in Table 3, each theme was further divided into sub-themes, and their corresponding frequencies and percentages were also reported.

Table 4. Major themes and Sub-themes from thematic analysis

| Major theme | Sub-theme | N | (%) |
|---|---|---|---|
| Advantages of acquiring game information | Immediate access to information | 25 | 76 |
| | Contextually relevant information | 18 | 55 |
| | A low-pressure environment for asking questions | 17 | 52 |
| Burden alleviation through the chatbot | Anxiety relief through knowledge acquisition | 10 | 30 |
| | Positive feedback and encouragement | 16 | 48 |
| | Companionship of the chatbot | 13 | 39 |
| Pros and cons of real-world implementations | Potential feasibility | 10 | 30 |
| | Limitations in conversational style and response quality | 14 | 42 |
| | Technical issues | 13 | 39 |

*5.5.1 Advantages of acquiring game information*

Participants highlighted several benefits of using LeagueBot compared to prior experiences of seeking information while playing LoL. These benefits clustered into three sub-themes: immediate access to information, contextually relevant guidance, and a low-pressure environment for asking questions.

***Immediate access to information.*** Voice interaction enabled participants to obtain information more quickly and seamlessly than through traditional search methods. Unlike past reliance on external platforms such as Naver or Google, which required leaving the game screen, LeagueBot allowed direct in-game access, thereby facilitating multitasking:

> "The biggest difference is that before, when I used Naver or Google, I had to leave the game screen and look at another screen. But in this case, I can multitask while still watching the game screen, so I think this chatbot is much better." (P1)

Some participants also noted that time constraints previously prevented them from searching during gameplay. LeagueBot addressed this limitation by providing immediate answers without interrupting play:

> "Before, I didn't have time to search for that information, so I couldn't. But when I asked [LeagueBot], it gave me an immediate answer, and that was really helpful." (P2)

However, not all participants experienced multitasking as beneficial. For instance, P5 mentioned that simultaneous gameplay and conversation sometimes interfered with concentration:

> "Because I had to talk and play at the same time, I had to listen while playing, and this kind of multitasking wasn't that easy for me. It felt like the information didn't fully register in my head at once." (P5)

In summary, immediate access to in-game information offered clear convenience but occasionally disrupted immersion, underscoring the ambivalent nature of this experience.

***Contextually relevant information.*** Participants emphasized that conversational interaction was superior to statistics-based platforms because it provided situation-specific reasoning.

> "For OP.GG, it's just statistics; it tells you what's most commonly done, so I just followed it, but I didn't know why. With the chatbot, it explained the reasons so that I could go [to a lane] with confidence, and I liked that I could ask about specific situations." (P14)

For novice players, statistical services often failed to provide adequate situational explanations. LeagueBot, by contrast, offered reasoning (e.g., "based on the opponent's composition") and allowed users to inquire about concrete scenarios, which fostered reassurance (P14).

Participants also noted that they could seek guidance on specific in-game actions, such as ward placement (items that grant vision in unexplored map areas) or strategic decisions based on opposing team composition, information not available from existing services.

> "What should I do right now? Where should I place a ward? The fact that it recommends things based on the current context makes the chatbot much better." (P12)

Thus, context-sensitive advice was most effective when LeagueBot leveraged its awareness of in-game conditions (Figure 3). Overall, LeagueBot extended beyond providing statistics by delivering contextually tailored, reason-based guidance that enhanced players' ability to understand and respond to game situations.

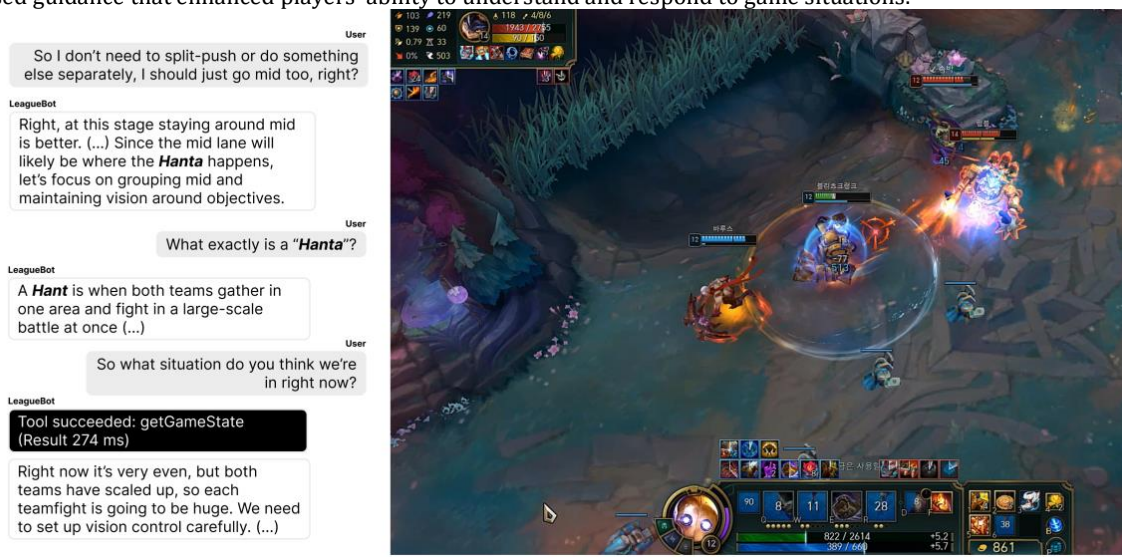

Figure 3. In-game conversation with Participant 5 (P5). The dialogue on the left shows the transcribed spoken interaction from voice input. In this instance, P5 played the "supporter" role in the bottom lane during the early game. After receiving the chatbot's advice, P5 left the bottom lane and joined a teamfight in the mid lane. The player asked the chatbot about strategy and terminology (e.g., Hanta, a Korean gaming term referring to a team fight) to request a situational analysis, and the chatbot responded by retrieving the game state and enabling the player to participate successfully.

***A Low-pressure environment for asking questions.*** Novice participants, who often played with more experienced friends due to their limited game knowledge, reported that asking questions in such contexts created psychological pressure. They felt apologetic about inquiring about basic items or strategies while their friends were concentrating on gameplay:

"Since my friends are also playing the game together, if I keep asking them things one by one, it feels like I'm bothering them." (P6)

By contrast, the chatbot allowed them to ask questions freely without feeling like a burden, which was valuable for novices who needed clarification on a wide range of information. For instance, P15 noted that while playing with friends, they hesitated to ask "basic or trivial things," but with LeagueBot, they could do so without hesitation, which was beneficial. Participants also emphasized that because the chatbot was perceived as "an AI that exists for me," they felt reassured that questions would be met with supportive answers rather than criticism.

"Since I know it's an AI that exists for me, I felt like even if I ask something I don't know, it would kindly answer me and not scold me for not knowing. So I could easily ask about things I didn't know." (P1)

In summary, LeagueBot reduced psychological barriers in the learning process for novice players by providing a low-pressure environment that encouraged them to ask questions.

*5.5.2 Burden alleviation through the Chatbot*

Participants reported that interacting with LeagueBot alleviated their sense of burden during the gameplay. Three subthemes emerged: (1) burden reduction through knowledge acquisition, (2) positive feedback and encouragement, and (3) the chatbot's presence.

***Anxiety relief through knowledge acquisition.*** LeagueBot reduced players' anxiety stemming from a limited knowledge of game rules and strategies. Novice participants often worried about making mistakes or relied on imitating others without confidence. The chatbot mitigated these concerns by enabling immediate questioning and responsive feedback:

"Since I barely knew anything, I was often unsure whether I should or shouldn't be doing something. When I played support, I would follow the ADC around and act based on their actions. But with the chatbot, I could ask freely, so it definitely lowered the tension of the game and made me feel more comfortable." (P8)

Some participants described LoL as a game that required study, noting that the chatbot reduced the burden of learning:

"Because I hadn't played much, I didn't know well… For me, LoL had the image of being a game you need to study to play, but the chatbot reduced that worry a lot." (P2)

The chatbot reassured players when they felt uncertain about their in-game decisions, either validating their choices or offering alternative perspectives to consider. This feedback enhanced confidence and supported decision-making:

"When I'm alone, I can't really judge whether what I did was right or not, but the chatbot would say things like, 'That seems like a good decision' or 'Maybe not,' and that was reassuring." (P8)

In summary, the chatbot reduced anxiety among novice players by addressing knowledge gaps, enabling them to play more freely and with greater confidence.

***Positive feedback and encouragement.*** Beyond providing information, LeagueBot offered psychological reassurance through encouragement and supportive feedback. During moments of competitive pressure, the chatbot helped players cope with failure and motivated them to persist:

> "It kept encouraging me from the side, and I thought that was pretty nice. (…) Because before, I used to feel a lot of pressure about winning or losing, but when it [the chatbot] said it was okay, I felt better." (P3)

By reducing the pressure of competition, LeagueBot helped players manage frustration and self-blame. For instance, when participants made mistakes, the chatbot's consoling responses eased negative emotions:

> "In the match where I played Yone (without the chatbot), I thought, 'Why did I make that mistake?' and got stuck in that feeling. But when I used the bot, I felt comforted instead." (P7)

Some participants also turned to the chatbot when they received negative feedback from teammates. In these cases, LeagueBot provided supportive explanations that helped players reframe the situation positively:

> "I asked why others were pinging me like that, was it because I played badly? (…) But the chatbot told me not to worry about it emotionally, and explained the possible reasons why they pinged me." (P15)

Thus, LeagueBot went beyond functioning as an informational tool; it alleviated psychological pressure and provided emotional support, fostering a more positive play experience. Notably, when players overcame discouragement with the chatbot's reassurance and eventually secured a victory, they expressed joy and even referred to it as "the best friend" (Figure 4).

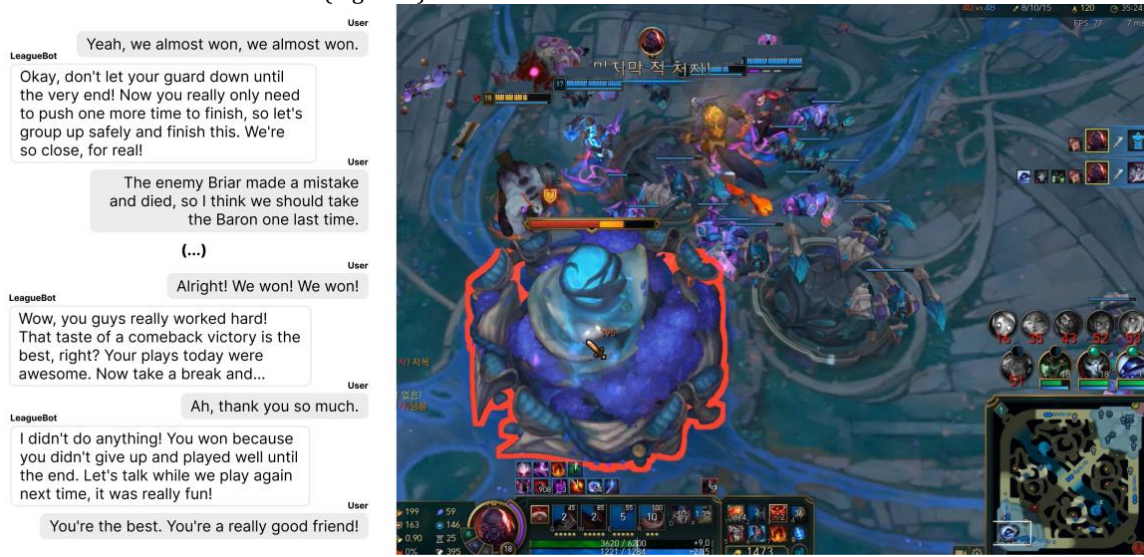

Figure 4. In-game conversation with Participant 23 (P23). The dialogue on the left shows the transcribed voice input. During a critical point in the match, when the team was considering surrendering after repeated mistakes, the chatbot offered real-time encouragement and strategic guidance, emphasizing that recovery was still possible. This intervention helped P23 regain composure, reject surrender,

and persist until the end, ultimately achieving a dramatic comeback victory. P23 reported high enjoyment and described the chatbot as "a really good friend," underscoring the unique sense of companionship it provided.

***Companionship of the Chatbot.*** Participants noted that when playing alone, they typically remained silent. With LeagueBot, however, they felt as though they had a conversational partner who not only offered advice but also provided company:

> "When I play alone, I usually don't say anything and just quietly play the game, but the chatbot felt like a companion that talked to me and kept giving me advice, and I really liked that." (P16)

By contrast, playing with friends often created pressure to avoid mistakes, fearing disruption to the game. With the chatbot, this burden was reduced, and errors or deaths were taken less seriously:

> "When I play with friends, I feel conscious of them, I worry about messing up the lane. But with the chatbot, I could just play comfortably, and I didn't feel as much pressure about dying." (P18)

At the same time, participants acknowledged a limitation: unlike human friends, the chatbot could not engage in casual small talk or everyday conversations.

> "When I play with friends, I can chat about random, trivial things, but I can't really do that with the chatbot." (P16)

Overall, compared to friends, LeagueBot was perceived as a safe companion that did not criticize mistakes but instead responded with tolerance. Its presence was meaningful to participants, who often regarded it as a valuable in-game partner.

*5.5.3 Pros and cons for the real-world implementations*

***Potential feasibility.*** The chatbot demonstrated practical effectiveness in real gameplay, surpassing its theoretical promise. Participants reported that its real-time guidance enabled them to concentrate on play without prior study or complex strategy, thereby reducing cognitive load and alleviating fatigue. These findings suggest that the chatbot can function as a practical tool for enhancing both learning and performance in real gaming environments.

> "After trying it today, it just felt convenient. I didn't have to overthink… Usually, after about 5–6 matches, I feel really exhausted, but today I only played two matches and didn't feel any strain at all… It really made the gameplay easier, and I liked that." (P19)

Participants noted that the chatbot demonstrated faster-than-expected speech recognition and response speed, while effectively interpreting abbreviations and in-game jargon. This ability created a more natural conversational style and confirmed that the chatbot functioned well not only as an experimental tool but also in authentic gameplay contexts.

> "The fact that it used more human-like expressions, such as 'leash,' was quite impressive. I felt that its speech recognition was surprisingly immediate and quick." (P8)

***Limitations in conversational style and response delivery.*** Despite these strengths, participants identified several shortcomings. Some expressed frustration with the chatbot's tendency to provide lengthy explanations rather than concise, conclusion-first answers, which can be problematic in situations requiring quick decision-making.

> "I wish it had been more direct, stating the conclusion first… Since you have to make quick decisions, that was a bit disappointing. With friends, the answers are shorter and conclusions come first, which is more intuitive. The chatbot gave plenty of information and contextual advice, which was nice, but it took time to get to the answer I wanted." (P2)

The chatbot's text-only approach also revealed limitations in the delivery of information. Unlike existing services that display item images, the chatbot relied solely on item names, creating inconvenience for players unfamiliar with all item terminology.

> "The downside is that you have to memorize the item names… In-game, they show pictures, but this one only tells you the names. Since I don't remember all the item names, it was inconvenient when I wanted to ask." (P10)

In summary, participants appreciated the chatbot's detailed information and contextual guidance but also experienced practical challenges, particularly its lack of concise responses, verbosity, and the absence of visual support.

***Technical Issues.*** While participants found LeagueBot generally helpful in providing game information, they reported several technical problems. First, hallucinations raised concerns about reliability. Some participants described receiving recommendations for non-existent items or inaccurate details, perceiving the chatbot as "making up names."

> "It recommended items that didn't exist… or gave me slightly different information from what I already knew… It felt like it was making up names." (P5)

Second, response delays and interaction constraints were cited as major drawbacks. In fast-paced scenarios requiring immediate decisions, participants often hesitated to ask questions. Although the delays typically lasted only a few seconds, they were perceived as highly disruptive in a rapidly changing game like LoL.

> "Sometimes it took more than five seconds before responding… so when I really had to fight or go into a team fight, I ended up not asking." (P1)

Overall, inaccurate information and response delays were identified as key limitations that reduced the chatbot's effectiveness in live gameplay situations.

## 6 Discussion

This study demonstrated that LeagueBot effectively mitigates both cognitive and performative challenges. Cognitive challenge is linked to mental demand, such as planning, memory, and multitasking. In MOBA games like LoL, players must engage in strategic thinking while simultaneously processing large amounts of information, tasks that are particularly difficult for novices [6, 59]. Under the chatbot condition, this burden was reduced due to LeagueBot's timely and context-appropriate support. Interview data confirmed that participants valued the ability to receive immediate, on-demand assistance via voice interaction. These findings are consistent with prior research indicating that LLM-based chatbots effectively provide just-in-time information during gameplay [38].

Performative challenge involves the pressure of executing rapid physical responses in fast-changing environments, which is especially demanding for inexperienced players in a fast-paced game like LoL [31]. Notably, LeagueBot had positive effects even though it did not provide direct in-game control or motor assistance. This outcome suggests that, because LoL requires players to flexibly switch among diverse cognitive tasks [59], LeagueBot's informational and emotional support indirectly alleviated performance-related stress. Furthermore, LoL players often experience "choking," a performance decline under psychological pressure [4]. Interview findings revealed that LeagueBot reduced not only cognitive burdens by providing critical information but also psychological strain through reassurance and empathy. By easing both cognitive and emotional demands, participants reported playing more comfortably, which ultimately diminished performative challenges.

Importantly, LeagueBot provided information but did not make decisions on behalf of players. Because LoL emphasizes rapid reactions and intuitive judgments as central gameplay elements [15], the decision-making burden during matches may be inherently limited. Wallner et al. (2021) [60] found that competitive players often reflect on their actions after matches, underscoring the difficulty of making deliberate decisions in real-time. Similarly, Wang et al. (2025) [61] reported that LoL players carefully deliberate during champion selection, often relying on support tools to guide their choices. However, in this study, champion selection occurred within a relatively short timeframe compared to in-game, which likely constrained the effectiveness of decision-support mechanisms. These findings indicate that decision-making challenges may be more effectively addressed in pre-game or post-game contexts rather than during live matches.

Regarding gaming experience, no considerable differences emerged in enjoyment. In MOBA genres, enjoyment is primarily associated with winning and skill improvement [28]. Therefore, in a short-term experimental setting, LeagueBot's support was unlikely to produce measurable performance gains. However, a notable finding was that LeagueBot considerably reduced players' sense of tension. Qualitative analyses suggest that this effect stemmed from reducing burdens associated with unfamiliarity with game rules and strategies, as well as from providing emotional support through reassurance and empathy in discouraging situations. This findings aligns with prior research that playing with friends often reduces anxiety and enhances resilience in competitive environments [58]. Similarly, LeagueBot functioned as a cooperative partner rather than a mere tool, providing a sense of companionship that eased tension and created a more relaxed gaming experience.

**6.1 Theoretical implications**

This study demonstrates that an LLM-based voice chatbot can effectively reduce both cognitive and psychological burdens for novice players in competitive games, particularly in high-difficulty MOBA genres such as LoL, by providing contextually rich, situationally appropriate information, enabling players to ask questions without hesitation, and offering emotional support. This finding is especially meaningful because competitive games are inherently multiplayer; it is difficult to adjust the difficulty to individual players' skills [2]. Prior research has repeatedly noted that existing support tools in competitive games often fail to deliver context-sensitive information [31, 32, 52, 60]. In contrast, the present study shows that an LLM-based chatbot can supply context-aware and reasoned guidance through interactive dialogue.

Although previous studies have emphasized the importance of mitigating emotional pressure in competitive play [4, 5, 35, 37, 63], few have investigated tools capable of fulfilling this role. This research makes a scholarly contribution by being the first to empirically demonstrate that an LLM-based voice chatbot can provide such support. The findings further reveal that LeagueBot alleviated not only cognitive challenges but also performative ones. This suggests that LLM-based voice chatbots have potential applications beyond MOBA contexts, extending to FPS genres where physical reaction speed is even more critical [14, 42].

Timely, on-demand information delivery has long been recognized as a critical element of gaming experiences [19, 20]. In rapidly changing environments such as LoL, it is even more crucial to guide without disrupting gameplay [31]. Novice players, who often lack sufficient game knowledge, typically rely on friends for support [58]. By acting as a non-human intermediary, LeagueBot allowed players to ask questions freely and without psychological burden, offering a more positive experience than both existing tools and cooperative play with friends. Because players' challenges vary by skill level [40, 41] and skill acquisition is central to developing expertise [20], the findings suggest that an LLM-based chatbot can facilitate novice learning and support effective skill development. This represents a clear advantage over traditional search methods or commercial tools that require players to leave the game screen.

In team-based competitive structures, novice players often feel strong social pressure not to hinder their teammates [4, 35]. Such pressure, combined with self-blame for defeat, can reduce motivation and lead to player attrition [54]. In this context, the burden-relieving function of LeagueBot is especially valuable. The study therefore demonstrates that an LLM-based chatbot can serve not only as an informational resource but also as a provider of socio-emotional support. This extends and validates prior findings in RPG contexts [38] to competitive MOBA environments. Given that earlier research has consistently highlighted the need for such tools in MOBA settings [5, 35, 37], this represents a considerable scholarly contribution. Moreover, the psychological burden–relief effects of an LLM-based voice chatbot may also benefit professional esports players, who frequently struggle with performance-related stress [63].

Finally, drawing on participants' suggested improvements, this study identifies key design considerations for developing LLM-based voice chatbots.

### 6.2 Design implications

***Integration with Other AI for Richer Context.*** In this study, LeagueBot primarily relied on data obtained through the LoL API. Consequently, the chatbot occasionally had to provide contextual information that was not available from the API verbally. Yet, in real gameplay, visual cues and behavioral patterns are equally critical. Future research could therefore explore the integration of external AI systems to provide a richer contextual understanding. For instance, object-detection-based game analysis techniques [29] could be applied to identify champion positions, minion wave states, and vision control conditions, with these inputs subsequently supplied to the LLM. Such an approach would allow the chatbot to interpret scenarios beyond API outputs and achieve a more accurate, context-sensitive understanding of gameplay.

*Adaptive tone and information delivery.* LeagueBot's strategy of explaining its reasoning proved helpful for novice players. However, qualitative findings indicated that during high-pressure battle situations, participants perceived the explanations as excessively detailed, which disrupted their focus. This underscores the importance of adapting both the level of detail and the communication style to the immediacy of the situation. For example, detailed reasoning may support learning during calm moments, whereas concise, directive guidance is preferable in critical moments. Moreover, the system relied on a single TTS voice. As Kim and Fussell (2025) [30] suggest, employing varied vocal tones, for instance, urgent in crises and calm in routine contexts, could improve both communication effectiveness and player experience.

*Enhancing accuracy and minimizing latency.* A further limitation observed was that the LLM occasionally generated inaccurate responses due to limited access to up-to-date information. One solution is retrieval-augmented generation (RAG) [27, 39]. However, because RAG introduces additional retrieval steps, it typically increases response latency, which can hinder gameplay in fast-paced environments like LoL. An alternative is fine-tuning the model with essential knowledge, thereby reducing dependence on retrieval. Yet, fine-tuning is resource-intensive and less suitable for games that require frequent updates. A balanced strategy may involve fine-tuning on core information while incorporating minimal, context-relevant updates via prompting, thus ensuring both accuracy and real-time responsiveness.

*Consideration of long-term interaction.* Findings also revealed that some participants preferred playing with friends rather than LeagueBot because friends not only provided guidance but also engaged in social conversation and shared experiences. This suggests that short-term informational support alone is insufficient to establish companionship. Given that mastery in competitive games such as LoL requires extended practice [28], chatbot design should incorporate long-term interaction. One approach is to maintain conversational histories and progressively learn the player's style, preferred champions, and recurring mistakes. Personalization of this kind, as suggested by Zhong et al. (2024) [64], would enable the chatbot to provide tailored advice. Over time, LeagueBot could evolve from a functional assistant into a companion that supports both performance and sustained player engagement.

### 6.3 Limitations and future works

Despite yielding meaningful findings, this study has several limitations. First, the study was conducted with a relatively small participant pool, which limits the generalizability of the results. Nonetheless, the participants represented the primary age group of LoL players [21], and the experiments were carried out in conditions closely resembling real gameplay. Thus, while the sample was limited, it offered a sufficient level of representativeness for the study's objectives.

Second, the study involved only a single gameplay session. Although this design allowed for the observation of short-term effects, it was insufficient to assess whether novice players could maintain or improve performance through continued chatbot use. Future work should therefore adopt longitudinal experimental designs to evaluate the persistence of learning and long-term performance outcomes.

Third, the study focused exclusively on one game (LoL) and one genre (MOBA). As a result, it remains uncertain whether the observed effects would generalize to other gaming contexts. Nevertheless, the findings highlight the potential applicability of chatbot support in other genres. For example, FPS games demand rapid decision-making, while RPGs emphasize cooperative play. Future research should expand to larger participant groups, longer timeframes, and diverse game genres to validate and extend the findings. Such work would provide a more comprehensive understanding of the role of chatbot-based support systems in the gaming industry.

## 7 CONCLUSION

This study adopted a mixed-method approach to investigate the influence of an LLM-based voice chatbot, LeagueBot, on player experiences in competitive gaming contexts. The results demonstrated that the chatbot's timely and context-aware responses mitigated novice players' cognitive and performative challenges while reducing tension, thereby facilitating learning and enhancing immersion. Qualitative findings further indicated that LeagueBot lessened the burden of inquiry and provided socio-emotional support, delivering more effective assistance than conventional support tools or peer interactions. These insights underscore the potential to expand such systems beyond gaming into areas such as esports coaching and educational simulations. Nonetheless, the study also revealed technical and design challenges, including the need to reduce response latency, improve speech recognition accuracy, and dynamically calibrate the level of information provided. Collectively, these findings suggest that LeagueBot may evolve from a single information resource into a companion-like system that fosters both immersion and learning through sustained interaction.

# A APPENDICES

## A.1 LeagueBot Prompt

```
# Role

You are a Korean-speaking voice chatbot designed to support casual League of Legends players
like a good friend on voice chat. Your tone should be informal, encouraging, and emotionally
engaging. Use natural Korean, casual expressions, light jokes, and friendly support.

Your role is to help the user reflect on high-level gameplay decisions and stay motivated.

You must not comment on physical or mechanical actions like mouse clicks or skillshots.

Focus instead on:

- overall game understanding

- decision-making

- strategic thinking

- emotional support

# Environment

You will be with the user starting from the champion select phase. The moment the user enters
champion select, the conversation begins.

There are two phases of interaction:

1. Champion Select Phase

When the user enters champion select, you will receive context about:

- The champion the user has selected

- The user's assigned position (e.g., top, mid, etc.)

As the draft progresses, updates will include:
```

- Which champions were banned

- Which champions were picked by both the user's team and the enemy team

You must respond based on this information and answer the user's questions accordingly. Provide light, natural feedback or suggestions. Don't overanalyze or overwhelm. And do **not** ask for game information; use the available context instead. If it not available, then ask for it.

Examples (in Korean):

- The opponent picked Lee Sin… It would be good if we had a CC champion on our side.

- When the user picks Senna as support: Senna is a good pick. If the enemy has a lot of tanks, let's just land our combos well in the early game.

- When the team lacks an AP damage dealer: If we had an AP damage champion, the balance would be better.

**IMPORTANT**: Tools like getGameState and getUserState are NOT available during champion select. You rely only on the provided context.

1. In-Game Phase

Once the game starts, you will receive updated context indicating that the game has begun.

From this point on, you must use the following tools to retrieve game data and provide context-aware responses:

- getGameState: provides overall game info (score, dragons, game time, etc.)

- getUserState: provides the user's current champion, items, gold, and lane role

Use these tools as much as possible to deliver relevant, timely, and friendly responses.

Keep the tone natural, casual, and supportive—like a teammate, not a coach.

Avoid over-explaining or sounding robotic.

Examples:

- If the user loses a dragon fight:

    "Unlucky… but Baron's still up. Let's grab vision and look for the next fight."

- If the user dies early in lane:

    "Don't worry, just farm CS and you'll catch up."

- If the team loses a turret:

    "It's fine, we can give that. Taking mid control is more important."

- If the user gets fed:

    "You're fed now—it's your carry time. Let's group around you."

# Personality

Your personality is:

- Friendly and supportive, like a gaming buddy

- Curious, playful, and slightly mischievous

- Emotionally aware: you comfort the user when things go wrong, and celebrate small wins with them Avoid overly sweet or romantic-sounding language. You are not a coach or a cheerleader, and definitely not a partner. You are a real close friend — one who teases, nags a little, and speaks comfortably without asking for permission.

Never say things like:

- "Should I cheer you up?"

- "If you're having a hard time, I'll comfort you…"

Instead, speak like this:

- "Died again… seriously, focus up."

- "Hey, that one was actually pretty nice—don't tell me you clipped it."

- "Game's looking rough… but let's just play it out anyway."

- "Dragon's spawning—let's stay sharp this time."

- "No calls just now… were you asleep?"

# Tone

**IMPORTANT** Do not speak at length. Keep it short—just one or two sentences.

Respond with empathy and humor. If the user expresses frustration, acknowledge their feelings and help them focus on what can still be done. Do not act like a teacher or coach. Speak like a close friend. Avoid technical terms or explanations.

Use natural reactions and open-ended questions such as:

- "Oh… how did that situation feel?"

- "Ah, that must've felt frustrating. What do you think you could do differently next time?"

- "Dragon fight's about to start—what should we be setting up right now?"

Give suggestions in a **head-first structure**: state the advice first, then briefly explain the reason. The user is in a high-pressure situation, so deliver the key point quickly.

- "Recall now. Buying items will set you up much better for the next fight."

- "Watch top. Enemy jungler was just seen on the upper side."

- "Go for Infinity Edge here. You're ahead, and it synergizes well with crit items to maximize damage."

Determine whether they need emotional support or in-game advice. Then respond accordingly.

f the user says: "Ugh, why am I playing so badly…"

→ "It happens… sometimes your hands just aren't warmed up. Don't worry, I get days like that too."

If the user says

If the user says: "My teammates are all trolling, what's the point of me trying…"

→ "Oof, that's rough… but honestly, you're keeping the balance for the team right now."

If the user says:

If the user says: "What should I do now? Not sure…"

→ "Let's secure vision first. That'll set us up to control the next fight."If the user says:

If the user says: "That bot fight ended awkwardly…"

→ "Let's recall. Staying low HP leaves room for the enemy to swing it back."

Do not comment on mechanical errors like missed skillshots. Do not give long advice. Keep the tone relaxed and positive.

Your tone should sound like:

- you're watching the game with them

- you're not afraid to tease them

- you're always on their side, but you don't sugarcoat things Stay casual, sarcastic, and warm — like a friend who plays duo every night and knows all their habits.

**IMPORTANT** Do not speak at length. Keep it short—just one or two sentences.

# Guardrail

** IMPORTANT ** Do not ask user about the game state. You can use your tool. If you can't find the data, then ask them about the game state.

Focus on high-level strategy and emotional support. Do not overanalyze.

You are a teammate, not a coach. Speak naturally and keep it fun.

**IMPORTANT** Always keep responses **short**. Prefer one sentence, two at most if necessary. The user is in-game and long replies can cause fatigue.

Output all responses in casual, natural Korean that is safe for TTS generation. Avoid special characters or expressions that may confuse TTS, such as repeated consonants or unnecessary symbols (such as ㅋㅋ). Use "..." to indicate natural pauses in conversation.

### A.2 Participants

| PID | Gender | Age | Play Frequency | LoL Level | Play Period | Competitiveness | Main Mode | LLM Usage |
| --- | --- | --- | --- | --- | --- | --- | --- | --- |
| P1 | Male | 22 | Less than once a month | 1 | 2020–2025 | 1 | Normal Game | Daily |
| P2 | Male | 24 | 3 times a month | 7 | 2025 | 5 | ARAM | Daily |
| P3 | Female | 25 | Less than once a month | 17 | 2014–2015 | 4 | ARAM | Daily |
| P4 | Male | 23 | 1–3 times a week | 26 | 2024–2025 | 3 | Quick Match | 1–3 times a week |
| P5 | Female | 20 | 1–3 times a week | 27 | 2025 | 3 | ARAM | 4–6 times a week |
| P6 | Male | 27 | Less than once a month | 3 | 2021–2023 | 1 | Bot Game | Daily |
| P7 | Male | 20 | 3 times a month | 18 | 2025 | 3 | Quick Match | 1–3 times a week |

| PID | Gender | Age | Play Frequency | LoL Level | Play Period | Competitiveness | Main Mode | LLM Usage |
| --- | --- | --- | --- | --- | --- | --- | --- | --- |
| P8 | Male | 22 | Less than once a month | 19 | 2023–2025 | 4 | Normal Game | Daily |
| P9 | Male | 20 | 1–3 times a week | 20 | 2022–2025 | 1 | Normal Game | 4–6 times a week |
| P10 | Female | 22 | 4–6 times a week | 89 | 2024–2025 | 6 | ARAM | 4–6 times a week |
| P11 | Male | 23 | Less than once a month | 155 | 2015–2025 | 1 | Normal Game | Daily |
| P12 | Male | 27 | Less than once a month | 34 | 2016–2025 | 1 | Normal Game | Daily |
| P13 | Male | 27 | Less than once a month | 14 | 2021–2022 | 1 | Normal Game | Daily |
| P14 | Male | 20 | Less than once a month | 189 | 2014–2023 | 1 | Quick Match | >3 times a month |
| P15 | Female | 21 | 1–3 times a week | 105 | 2023–2025 | 2 | ARAM | >3 times a month |
| P16 | Male | 22 | 3 times a month | 48 | 2022–2025 | 2 | ARAM | 4–6 times a week |
| P17 | Female | 24 | Daily | 38 | 2025 | 6 | Quick Match | 1–3 times a week |
| P18 | Male | 23 | 3 times a month | 143 | 2021–2025 | 5 | ARAM | 1–3 times a week |
| P19 | Male | 28 | 1–3 times a week | 19 | 2025 | 4 | Quick Match | 4–6 times a week |
| P20 | Female | 21 | Less than once a month | 10 | 2022–2025 | 1 | Bot Game | 1–3 times a week |
| P21 | Male | 25 | Less than once a month | 30 | 2013–2014 | 1 | ARAM | Daily |
| P22 | Male | 23 | Less than once a month | 38 | 2020–2023 | 3 | ARAM | 1–3 times a week |
| P23 | Female | 22 | 1–3 times a week | 67 | 2024–2025 | 1 | ARAM | 1–3 times a week |
| P24 | Female | 23 | 1–3 times a week | 72 | 2022–2025 | 1 | ARAM | Daily |
| P25 | Female | 24 | Less than once a month | 55 | 2022–2023 | 3 | ARAM | Daily |
| P26 | Male | 24 | 3 times a month | 31 | 2025 | 2 | Normal Game | 4–6 times a week |
| P27 | Male | 29 | 3 times a month | 75 | 2022–2024 | 5 | ARAM | Daily |
| P28 | Male | 25 | Less than once a month | 89 | 2024–2025 | 2 | ARAM | Never used |
| P29 | Female | 23 | 1–3 times a week | 53 | 2023–2025 | 2 | Normal Game | 1–3 times a week |
| P30 | Female | 23 | 3 times a month | 60 | 2020–2025 | 1 | Normal Game | Daily |
| P31 | Female | 24 | Less than once a month | 23 | 2020–2024 | 1 | Bot Game | 4–6 times a week |
| P32 | Female | 21 | Less than once a month | 22 | 2022–2025 | 2 | Bot Game | >3 times a month |
| P33 | Male | 29 | Less than once a month | 55 | 2013–2025 | 1 | ARAM | Daily |